\documentclass[aoas,MSNbibl,nameyear,seceqn,dvips]{arximspdf}
\usepackage{graphicx}
\doi{10.1214/12-AOAS603} %kopijuoti is PTS
\volume{7}
\issue{2}
\pubyear{2013}
\firstpage{860}
\lastpage{882}

\makeatletter

\newcommand{\cal}{\mathcal}

\newcommand{\vecv}{{\mathrm{v}}}

\makeatother

\begin{document}
\begin{frontmatter}

\title{Visualizing genetic constraints}
\runtitle{Visualizing genetic constraints}

\begin{aug}
\author[A]{\fnms{Travis L.} \snm{Gaydos}\thanksref{t5,t6}\ead[label=e1]{travis.gaydos@gmail.com}},
\author[B]{\fnms{Nancy E.} \snm{Heckman}\corref{}\thanksref{t2}\ead[label=e2]{nancy@stat.ubc.ca}},
\author[C]{\fnms{Mark} \snm{Kirkpatrick}\thanksref{4}\ead[label=e3]{kirkp@mail.utexas.edu}},
\author[D]{\fnms{J.~R.} \snm{Stinchcombe}\thanksref{t2}\ead[label=e4]{John.stinchcombe@utoronto.ca}},
\author[E]{\fnms{Johanna} \snm{Schmitt}\thanksref{t3}\ead[label=e5]{jschmitt@ucdavis.edu}},
\author[F]{\fnms{Joel} \snm{Kingsolver}\thanksref{t6}\ead[label=e6]{jgking@bio.unc.edu}}\\
\and
\author[G]{\fnms{J. S.} \snm{Marron}\thanksref{t5,t7}\ead[label=e7]{marron@unc.edu}}
\runauthor{T. L. Gaydos et al.}
\affiliation{MITRE Corporation,
University of British Columbia,
University of Texas at Austin,
University of Toronto,
University of California,
University of North Carolina and
University of North Carolina}
\address[A]{T. Gaydos \\
MITRE Corporation \\
McLean, Virginia\\
USA\\
\printead{e1}}
\address[B]{N. E. Heckman \\
Statistics Department \\
University of British Columbia\hspace*{17.5pt}\\
Vancouver, BC\\
Canada\\
\printead{e2}}
\address[C]{M. Kirkpatrick \\
Section of Integrative Biology\\
University of Texas at Austin\\
Austin, Texas \\
USA\\
\printead{e3}}
\address[D]{J. R. Stinchcombe \\
Department of Ecology\\
\quad and Evolutionary Biology\\
University of Toronto\\
Toronto, Ontario\\
Canada \\
\printead{e4}}
\address[E]{J. Schmitt \\
Department of Evolution and Ecology \\
University of California\\
Davis, California \\
USA\\
\printead{e5}}
\address[F]{J. Kingsolver \\
Department of Biology \\
University of North Carolina\hspace*{22pt}\\
Chapel Hill, North Carolina\\
USA\\
\printead{e6}}
\address[G]{J. S. Marron \\
Department of Statistics\\
\quad and Operations Research\\
University of North Carolina\\
Chapel Hill, North Carolina \\
USA\\
\printead{e7}} %adresu isvedimo komanda gale!
\end{aug}

\thankstext{t5}{Supported by National Science Foundation EF-0328594.}

\thankstext{t6}{Supported by National Science Foundation IOS-061179.}

\thankstext{t2}{Supported by the National Science and Engineering
Research Council of Canada.}

\thankstext{4}{Supported by National Science Foundation Grant DEB-0819901.}

\thankstext{t3}{Supported by National Science Foundation Grant DEB-0129018.}

\thankstext{t7}{Supported by NSF Grants DMS-03-08331 and DMS-06-06577.}

% HISTORY:
\received{\smonth{3} \syear{2012}}
\revised{\smonth{8} \syear{2012}}

% ABSTRACT
%
\begin{abstract}
Principal Components Analysis (PCA) is a common way to study the
sources of variation in a high-dimensional data set. Typically, the
leading principal components are used to understand the variation in
the data or to reduce the dimension of the data for subsequent
analysis. The remaining principal components are ignored since they
explain little of the variation in the data. However, evolutionary
biologists gain important insights from these low variation
directions. Specifically, they are interested in directions of low
genetic variability that are biologically interpretable. These
directions are called \textit{genetic constraints} and indicate
directions in which a trait cannot evolve through selection. Here, we
propose studying the subspace spanned by low variance principal
components by determining vectors in this subspace that are simplest.
Our method and accompanying graphical displays enhance the biologist's
ability to visualize the subspace and identify interpretable directions
of low genetic variability that align with simple directions.
\end{abstract}

% KEYWORDS

\end{frontmatter}

%The following appeared long ago as a comment from Mark. Shall we
%include it here? Where/ I do not quite understand this.
%``As the number of traits (the number of phenotypic dimensions)
%increases, the potential for genetic constraints and the
%dimensionality of the null space may also increase %%(Dickerson
%1955).''

%s1 #&#
\section{Introduction} \label{sectIntro}

Evolutionary biologists study how the distribution of observable
characteristics of individuals in a population changes over
generations. These observable characteristics are called traits or
phenotypes and can be qualitative, such
as body color in a specific environment, or
quantitative. A quantitative phenotype can be a scalar such as
mass at a specified age, or a vector such as mass at several specified
ages, or a function such as mass
at a continuum of ages.

Changes in the distribution of traits can occur via many processes,
including mutation, selection and genetic drift (the change in the
distribution of genotypes that can occur in a finite population when
mating and reproduction are modeled as random processes). Here we
consider changes caused by selection. We consider changes within only
one generation. We characterize
changes by the expected change in phenotype,
and we assume that the population under selection is,
in essence, infinite.
The selection process determines which individuals in a population are
likely to produce viable offspring. Selection can occur naturally,
when, for instance,
small individuals are more vulnerable to predation, or artificially, as
in the selective breeding of
race horses. Selection causes the trait distribution of the
subpopulation of breeding individuals to differ from
that of the original population. This difference will persist into the
offspring population provided the trait has
some genetic component.

To understand the role of selection and genetics in evolution, consider
the following simple example. Suppose that,
in a population,
individuals taller than a certain height do not reproduce. Thus, the
breeding subpopulation
will have a smaller mean height than the original population. The
breeding parents' offspring also will have a smaller mean height provided
height has some genetic basis.
In this case, we say that selection on height leads to the evolution of height.

Thus, evolution requires both a selection process and a genetic
component. The selection
process must involve a trait with a genetic component. That genetic
component must differ between breeding and
nonbreeding individuals.

Clearly, genetic variation plays an important role in evolution. As we
will see, the amount of genetic variation actually determines the speed
at which selection causes evolutionary change. In nature, traits with
substantial genetic variation will respond rapidly, allowing the
species to adapt rapidly to changing conditions. Genetic variation is
likewise a critical variable for plants and animals that are used in
agriculture. Artificial selection (or selective breeding) has been used
for millennia to improve domesticated species, and it continues to be
one of the most important tools for increasing agricultural yield. The
amount of genetic variation present is one of the key criteria used by
animal and plant breeders to choose the traits for artificial
selection. In both natural and domesticated populations, traits with
little or no genetic variation are not able to respond much or at all
to selection. These traits are said to be \textit{genetically
constrained}, and these constraints play an important role in
determining how populations adapt [see \citet{Kirkpatrick1992}].

In this paper we propose methods to explore genetic constraints in
vector-valued traits.
The next section contains biology background, including a model for selection
and a characterization of genetic constraints as
eigenvectors of the genetic covariance matrix corresponding to zero eigenvalues.
Sections \ref{sectanalysis} and \ref{sectsimplicitybasis} describe our
proposed methodology for studying genetic constraints. Data analyses appear
in Section \ref{sectdata} and a simulation study in Section \ref
{sectsimulation}.

%s2 #&#
\section{Biology background}
\label{sectBio}

Biologists model an individual's quantitative trait in terms of
components, the simplest model
involving two components: a genetic component, $g$, inherited from
parents, and an environmental component, $e$,
such as availability of food. In this simple model, the true phenotype
is $g+e$ and
the observed phenotype, $y$, is
\[
y = g + e + \varepsilon,
\]
where $\varepsilon$ is additional sampling variation. We denote the
expected value of $g$
by $\mu$ and its variance/covariance by $\mathcal{G}$. If $g$ is
scalar, then $\mathcal{G}$ is its variance.
If $g$ is a vector of length $K$, then $\mathcal{G}$ is the $K$ by $K$
covariance matrix with $ij$th entry equal to the
covariance between the $j$th and $k$th component of $g$. If $g$ is a
function, say, if $g(t)$ is mass at age $t$, then
$\mathcal{G}$ is a bivariate function, with $\mathcal{G}(s,t)$ being
the covariance between mass at age $s$ and mass at age $t$.
The environmental effect
$e$ is a mean zero random component with variance/covariance $\mathcal
{E}$, with $\mathcal{E}$ defined in an analogous way
as $\mathcal{G}$.
The random components $g$, $e$ and $\varepsilon$ are defined so as to be
uncorrelated, so the covariance of the true phenotype is $\mathcal{G} +
\mathcal{E}$ and
the covariance of the observed phenotype is $\mathcal{G} + \mathcal{E}$
plus the variance/covariance of $\varepsilon$.
The marginal distributions of $g$ and $e$ are population and
generation dependent, while the marginal distribution
of $\varepsilon$ depends on the method of measuring the phenotypes.

The heritability of a scalar phenotype is the proportion of its
variance that is
attributable to genetics, that is,
the heritability is simply $h^2 = \mathcal{G}/(\mathcal{G}+ \mathcal
{E})$. Throughout, we assume that
$[\mathcal{G} + \mathcal{E}]^{-1}$ exists. To understand the role of
heritability in evolution,
consider our simple example where the selection mechanism prevents tall
individuals from producing offspring. First suppose that height has
zero heritability in the population,
that is, all variability in height is simply due to
environmental effects. Then, intuitively,
the distribution of heights in the next generation will be the same as
the distribution in
the original population,
provided both generations are raised in similar environments.
However, if height has nonzero heritability, that is, if the genetic
component of height varies
across individuals, then
the distribution of heights in the next generation will be different
from the
distribution in
the original population.
One would expect that the larger the heritability in the original population,
the bigger
the change in the distribution of heights in the next generation.

The mathematical theory that supports this reasoning, that
links heritability and evolution of a trait from one generation to the
next, is contained in the \textit{Breeder's equation}.
To define this equation, let $\mu_p$ be the expected phenotype in the
original population, $\mu_{p^*}$ the expected phenotype of the
reproducing adults,
and $\mu_o$ the expected phenotype of their offspring.
The Breeder's equation gives $\mu_o - \mu_p$, the expected response to
selection:
%
%e2.1 #&#
%
\begin{equation}
\label{Breeders} \mu_o - \mu_p = {\mathcal{G}}[
\mathcal{G} + \mathcal{E}]^{-1} \times(\mu_{p^*} -
\mu_p).
\end{equation}
In our height example, $\mu_{p^*}$ is less than $ \mu_p$ and so the
Breeder's equation tells us that the mean height in the
offspring population is less than or equal to that in the original population.
How much\vspace*{1pt} less depends on the value of the heritability ($h^2 ={\mathcal
{G}}[ \mathcal{G} + \mathcal{E}]^{-1}$)
and the strength of selection. The strength of selection determines
if particular individuals will reproduce. In our simple height example,
the strength of selection is determined by
the height cutoff for reproducing.
Thus, the strength of selection determines $\mu_{p^*} - \mu_p$.
Biologists define the \textit{selection differential} as $s = \mu
_{p^*} - \mu_p$.
The Breeder's equation also holds for multivariate phenotypes, where,
if the phenotype is a vector of
$K$ values, then $\mu_o, \mu_p$ and $ \mu_{p^*}$ are $K$-vectors and
$\mathcal{G}$ and $\mathcal{E}$ are $K \times K$ covariance matrices.
For a generalization of the Breeder's equation to function-valued
traits, see
\citet{KirkpatrickandHeckman}.

Biologists rewrite the Breeder's equation
in terms of the \textit{selection gradient}, denoted $\beta$.
The selection gradient is defined in terms
of a population's expected fitness, that is, its ability to reproduce,
under the specified
selection mechanism.
We can think of the selection gradient as the change in $\mu_p$
that selection appears to be making when ``choosing'' the breeding
individuals in the original population.
This is not, in general, equal to $s=\mu_{p^*} - \mu_p$,
the change that
actually occurs. To see the distinction between $\beta$ and $s$,
consider once again our simple height example, but suppose that the phenotype
is a vector in $\Re^2$ with components height and weight. Selection is
only acting
on height, not on weight, so the selection gradient's second component
is zero.
However, the second component in $s$ is not 0 since height and weight
are positively correlated: the selection on height means that both the
heights and weights
of the reproducing individuals will, on average, be smaller than those
in the original
population. One can show that the selection gradient $\beta$ and the
selection differential $s$
are related via the equation $s = [ \mathcal{G} + \mathcal{E}] \beta$.
This yields
an alternative expression for the expected response to selection in the
Breeder's equation in (\ref{Breeders}):
%
%e2.2 #&#
%
\begin{equation}
\label{Breedersbeta} \mu_o - \mu_p = {\mathcal{G}}
\beta.
\end{equation}
We consider the amount of genetic variation explained by the direction
of a unit vector $v$. This amount of variation is the magnitude
of ${\mathcal{G}}v$, that is, the magnitude of the expected response to
selection when $v$ is the selection gradient.

For more details on the Breeder's equation, the selection gradient and
the selection differential,
see Lande (\citeyear{Lande1976,Lande1979}),
\citet{Lande1983} or, for a statistician-friendly exposition,
\citet{Heckman2003}. For an extension of (\ref{Breedersbeta}) to
function-valued traits, see
\citet{GomulkiewiczandBeder} and \citet{BederandGomulkiewicz}.

From ({\ref{Breedersbeta}}), we can see the importance of an
eigenanalysis of $\mathcal{G}$
in understanding a population's ability to evolve under selection. The
magnitude of $\mu_o - \mu_p$
will be largest when the selection gradient, $\beta$, points in the
same direction as the leading
eigenvector of $\mathcal{G}$. The value of $\mu_o - \mu_p$ will be zero
if selection
acts in the direction corresponding to a zero eigenvalue of $\mathcal{G}$.
That is, the population's mean phenotype will not evolve if selection
acts in the direction of an eigenvector
of $\mathcal{G}$ corresponding to a zero eigenvalue. These directions
are called
\textit{genetic constraints}.
Eigenvectors corresponding to small but nonzero eigenvalues are also of
interest.
\citet{GomulkiewiczandHoule} provide tools to determine what
eigenvalues are considered small:
they model demography and evolution in a population experiencing
selection due to changing environmental conditions. They identify
critical levels of genetic variability, levels low enough to
effectively prevent the adaptive evolution that might result from selection.

%%%%%%%%%%%%%%%%%%%%%%%%%%%%%%%%%%%%%%%%%%%%%%%%
%s3 #&#
\section{Analysis of genetic variability}
\label{sectanalysis}

To better understand the directions of genetic variability, we
partition the sample space of $g$ into
two subspaces, the \textit{model space} and the \textit{nearly null space}.
The model space is a ``high
genetic variance'' subspace spanned by eigenvectors of ${\mathcal{G}}$
with large eigenvalues.
The nearly null space is the orthogonal
complementary ``low genetic variance'' subspace. Visualizing the nearly
null space provides information about
the existence and interpretation of genetic constraints.
This partitioning and the associated visualization tools were
introduced in \citet{Gaydosthesis}.

To explicitly define the model space and the nearly null space of a
covariance matrix $\mathcal{G}$,
let $\lambda_1 \ge\lambda_2 \ge\lambda_3 \ge\cdots$ be the
eigenvalues of $\mathcal{G}$,
and $v_1,v_2, v_3, \ldots$ the corresponding
orthonormal eigenvectors. We decide
which $\lambda_k$'s
to consider as large values, say, $\lambda_1,\ldots, \lambda_J$, and
assume that $\lambda_{J} $ is strictly greater than $\lambda_{J+1}$.
We define the model space
as the space spanned by $v_1,v_2,\ldots, v_J$ and the nearly null
space as the space
spanned by the remaining eigenvectors.
From the Breeder's equation~(\ref{Breedersbeta}),
we see that
$\mu_o - \mu_p$ is large if $\beta$ lies in the model space.
Specifically, for $\beta$ in the model space,
$\|\mu_o - \mu_p\|/\|\beta\| \ge\lambda_J$ and is largest when $\beta$
is a constant times $v_1$.
Conversely, if $\beta$ lies in
the nearly null space, then
$\|\mu_o - \mu_p\|$ will never exceed
$\lambda_{J+1}\|\beta\| $.

Interpreting the nearly null space is challenging since, typically,
eigenvectors corresponding to small eigenvalues are ``rough''
and may simply represent noise.
To study the nearly null space, we construct a new basis for this
space, ordered by
simplicity. If the simplest basis vectors are interpretable,
biologists can then study the possibility of genetic constraints. If
the simplest
basis vectors are not interpretable, then biologists might consider the
nearly null
space to represent noise.

Clearly, the choice of $J$ is important in the definition of the model
space and nearly null space.
One might carry out a sequence of hypothesis tests to choose~$J$, using
the procedures of
\citet{Amemiya1990}, \citet{AndersonandAmemiya} or
\citet{HineandBlows}. These authors consider testing for the
dimension of $\mathcal{G}$ when data
come from a half-sibling design, that is, where data are from
independent families and each family consists of half-siblings. Such data
can be modeled as an easy-to-analyze multivariate one-way
classification with random effects [\citet{WalshLynch}].
Hypothesis testing to determine $J$ in more complicated designs might
be challenging.
However, we do not recommend this hypothesis testing approach,
preferring instead an exploratory approach grounded in
the biology.
We recommend the usual techniques of calculating the proportion of
genetic variance explained, studying
scree plots and considering the interpretability of the associated
eigenvectors, combined with the calculations of critical levels as
defined in \citet{GomulkiewiczandHoule}.
\label{DickDavidcritical}
We also recommend that subject area specialists examine results for a
range of values of $J$, to
examine the interplay between proportion of variance explained and
biological interpretability of the resulting
model space and nearly null space.
These subject area opinions can provide a
more biologically meaningful and thus more compelling explanation of a
choice of reasonable values of $J$ than any test of significance or
other algorithmic approach.
In addition, studying a range of values of $J$ allows
the user to consider
small-scale and large-scale genetic variabilities.
\label{flip}

%%%%%%%%%%%%%%%%%%%%%%%%%%%%%%%%%%%%%%%%
%
%f1 #&#
%
\begin{figure}

\includegraphics{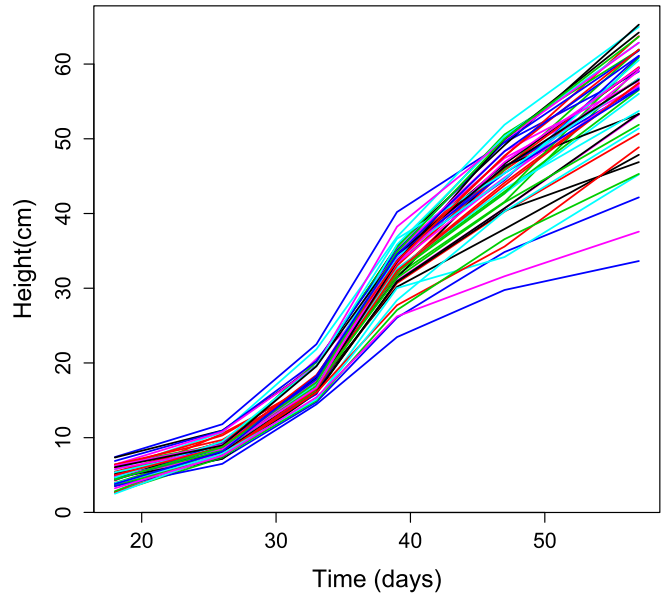}

\caption{Heights of 49 jewelweed plants raised in sun in a dense environment.
The heights are measured at six times and linearly interpolated. See
Stinchcombe et~al. (\citeyear{Stinchcombe2010}) for details of the experiment.}
\label{JWdatasun}
\end{figure}
%
%% create with matplot(xval,t(data),type="l",lty=1,xlab="Time
%(days)",ylab="Height(cm)",col="black",cex.lab=1.5,cex.axis=1.3,lwd=1.5)

%f2 #&#
%
\begin{figure}

\includegraphics{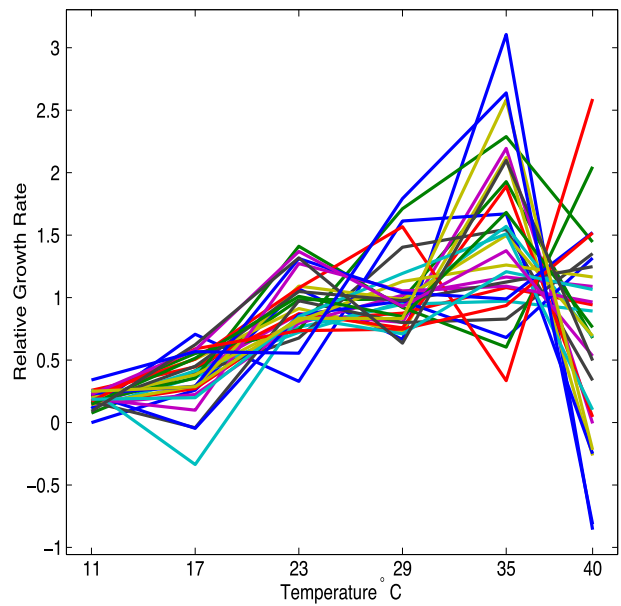}

\caption{Relative growth rates of the caterpillar \textup{Pieris rapae}
as a function of temperature.
The growth rates, in milligrams per hour, are measured at six
temperatures and linearly interpolated.
See Kingsolver, Ragland and
Shlichta (\citeyear{KRS2004}) for details of the experiment.}
\label{Catdata}
\end{figure}

%%%%%%%%%%%%%%%%%%%%%%%%%%%%%%%%%%%%%%%%

In summary, we use principal components analysis and simplicity
measures to define biologically interpretable directions of low genetic
variation, allowing biologists to
explore the possibility of the existence of genetic constraints.
We apply our method to two data sets, one of the heights
of jewelweed plants (\textit{Impatiens capensis}) measured at six ages,
the other of
growth rate measurements of the caterpillar \textit{Pieris rapae} at six
temperatures. The jewelweed data are described in
\citet{Stinchcombe2010}
and the caterpillar data in \citet{KRS2004}. The two data sets
are displayed in Figures \ref{JWdatasun} and \ref{Catdata}.
Descriptions of the data and the purpose of the experiments, along with
data analysis and discussion, are given in Section~\ref{sectdata}.

%s4 #&#
\section{Simplicity basis}
\label{sectsimplicitybasis}

A simplicity basis
for a linear subspace $\mathcal{V}$ of $\Re^K$ is an orthonormal basis
$\{w_1,\ldots,w_L\}$,
where the $w_k$'s are ordered according to some simplicity measure:
$w_1$ is the ``simplest'' element of unit length in
$\mathcal{V}$,
$w_2$ is the ``simplest'' unit-length element of ${\mathcal{V}}$ that
is orthonormal to $w_1$,
$w_3$ is the ``simplest'' unit-length element of ${\mathcal{V}}$ that
is orthonormal to $w_1$ and $w_2$,
and so forth. Such a
basis may help us to understand $\mathcal{V}$ since simple vectors are
usually the most
interpretable.

We consider quadratic simplicity measures,
that is, measures equal to $v' \Lambda v$. We assume throughout that
$\Lambda$
is a nonnegative definite symmetric matrix and, for interpretability,
that $\Lambda$ is defined so that
the simpler the vector the higher the simplicity score. If this is not
the case, that is, if $v' \Lambda v$
is small when $v$ is simple, we can instead use the simplicity measure
$v'(\lambda{\mathrm{I}} - \Lambda)v$, where
I is the identity matrix and
$\lambda$ is some number greater than or equal to the largest
eigenvalue of $\Lambda$.
Examples of quadratic simplicity
measures can be found in smoothing and penalized regression.
See, for instance, \citet{EilersMarx1996} or \citet
{SilvermanGreen1994}.
In the examples that follow, we think of the elements of a vector $v$
as evaluations of a function $f$:
$v = (\vecv_1,\ldots, \vecv_K)' = (f(t_1),\ldots, f(t_K))'$. In these examples,
we used a simplicity measure based on first divided differences:
\[
\sum_j \frac{(\vecv_j-\vecv_{j-1})^2}{ (t_j - t_{j-1})},
\]
which is a good approximation of $\int(f')^2$.
To transform this to a measure that is large for simple $v$'s, we use
the result of \citet{Schatzman2002} that
$\sum(\vecv_j -\vecv_{j-1})^2 \le4 \sum\vecv_j^2$ for all $v$'s. Our
simplicity measure is equal to
%
%e4.1 #&#
%
\begin{equation}
\label{eqfirstdivided} 4 v'v - \min_j
\{t_j-t_{j-1}\} \times\sum\frac{(\vecv_j-\vecv_{j-1})^2}{ (t_j -
t_{j-1})},
\end{equation}
which lies between 0 and 4 inclusive. The simplicity measure
(\ref{eqfirstdivided}) is just one of many possible smoothing-based
measures. Another good choice might be the measure used in cubic
smoothing spline regression, where a function $f$'s simplicity is
defined as $\int(f'')^2$, with a low value signifying simplicity. This
integral can be approximated using a Rieman sum of second divided
differences, yielding a quadratic form in $(f(t_1),\ldots, f(t_n))'$.

The simplicity basis of the subspace $\mathcal{V}$ for the simplicity
measure associated with a nonnegative definite
symmetric matrix $\Lambda$ is
easy to calculate. Let
$v_1,\ldots, v_L$ be an orthonormal basis of $\mathcal{V}$ and
let
$P$ be the $K\times L$ matrix with $k$th column equal to $v_k$.
So $P$ is a projection matrix onto ${\mathcal{V}}$. Let $\alpha_1,\ldots
, \alpha_L$ be the eigenvectors of
$P' \Lambda P$ with corresponding eigenvalues $\lambda_1 \ge\lambda_2
\ge\cdots\ge\lambda_L$. Then
it is straightforward to show that $\{ P \alpha_1,\ldots,
P \alpha_L\}$ is a simplicity basis of $\mathcal{V}$, ordered from most
simple to least simple.
When the eigenvalues are distinct, the basis is unique, not dependent
on the choice of $P$.
However, if, for example, $\lambda_1=\lambda_2 > \lambda_3$, then the
``simplest subspace'' of $\mathcal{V}$ is the span of $P \alpha_1$ and
$P\alpha_2$,
and this subspace does not depend on the $P$ that we choose.

%%%%%%%%%%%%%%%%%%%%%%%%%%%%%%%%%%%%%%%%%%%%%%%%%%%%%%%%

%s5 #&#
\section{Data analysis}
\label{sectdata}

For well-designed evolutionary biology studies such as those presented here,
the covariance matrix ${\mathcal{G}}$ is identifiable, estimable and
consistent. Typical methods of
estimation are via MANOVA, maximum likelihood or restricted maximum
likelihood (REML). See, for instance,
\citet{Searle2006} and \citet{WalshLynch}. These estimates
take into account the dependence in the data caused by individuals' relatedness.

For each data set, we carry out a principal components analysis of the
estimate of $\mathcal{G}$ and, for
all possible values of $J$,
we study the model space of dimension $J$ and the corresponding nearly
null space.
For our data sets, $J$ ranges from 0 to 6.
The supplementary material for this paper contains all seven plots of
the caterpillar analysis and all seven plots of the jewelweed
analysis. Here, we present just two of the seven plots for each data set.

%f3 #&#
%
\begin{figure}

\includegraphics{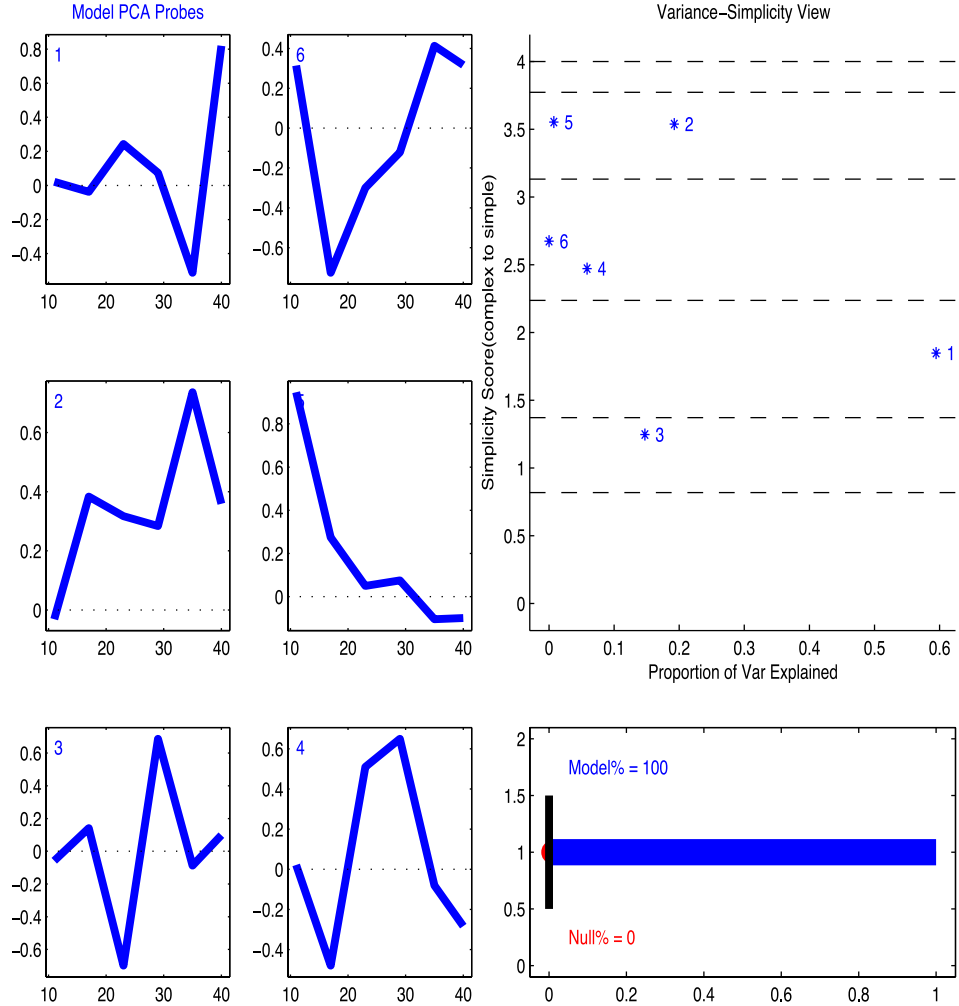}

\caption{Caterpillar data: PCA basis. The vectors
in the left panel are the six principal components vectors of the
estimated genetic covariance matrix. These vectors explain 100\% of
the genetic variance, as indicated in
the lower right plot.
The upper right plot shows each vector's simplicity score and the
percent of genetic variance
it explains.
The amounts of genetic variance explained by vectors~1 through 6 are,
respectively,
0.618, 0.200, 0.153, 0.061, 0.008, 0.}
\label{figcatpca}
\end{figure}

\begin{figure}

\includegraphics{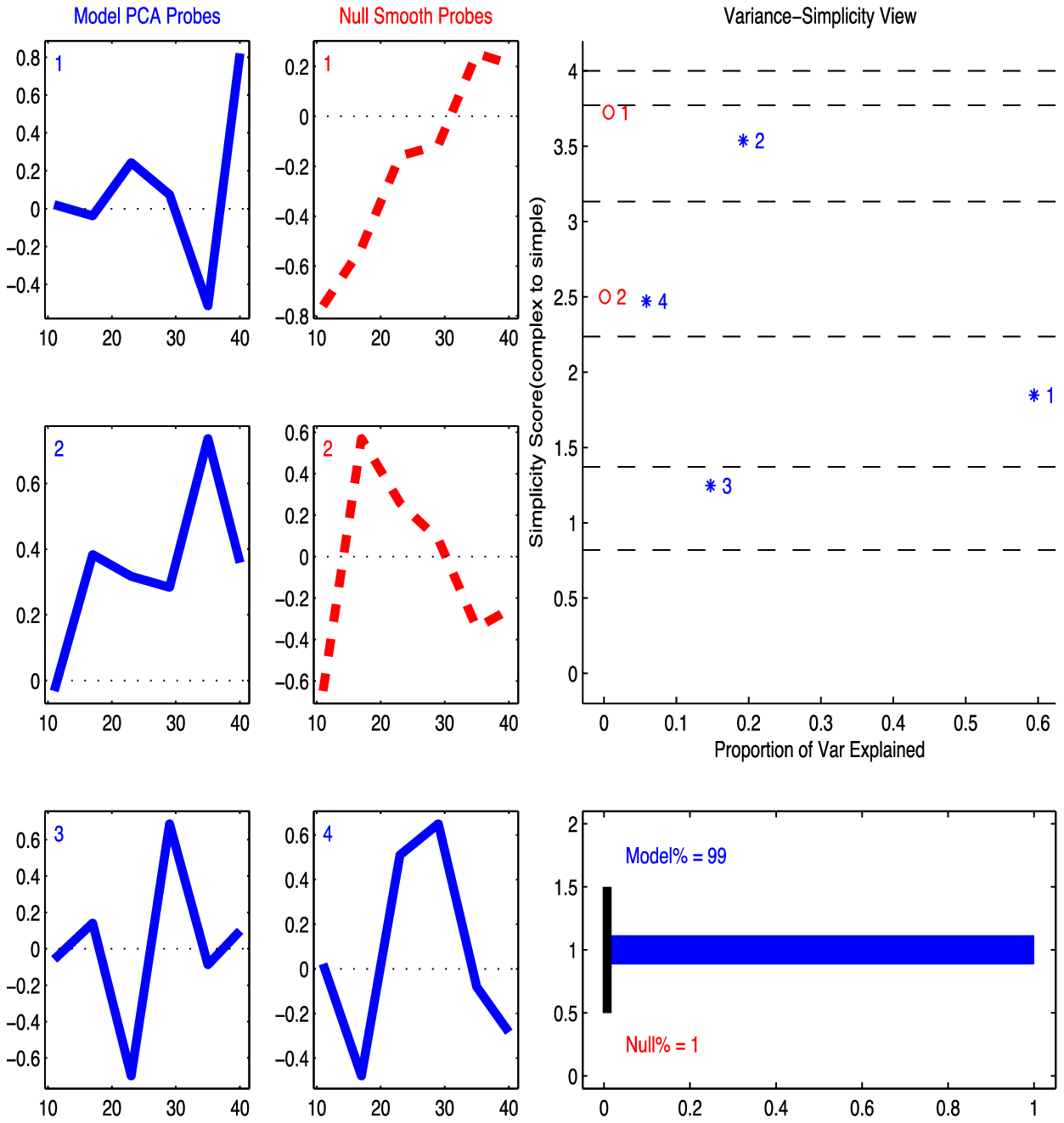}
 \caption{Caterpillar data: two-dimensional nearly null
space. The structure of the plot is as in
Figure~\protect\ref{figcatpca} except that the left-hand side shows the
PC basis for the four-dimensional model space (blue) and the simplicity
basis for the two-dimensional nearly null space (red dashed). The
simplest nearly null space vector is labeled with a red 1. The amount
of genetic variance explained by the simplest nearly null vector is
0.007 and by the second simplest 0.001.} \label{figcatsimplicity2}
\end{figure}

\begin{figure}

\includegraphics{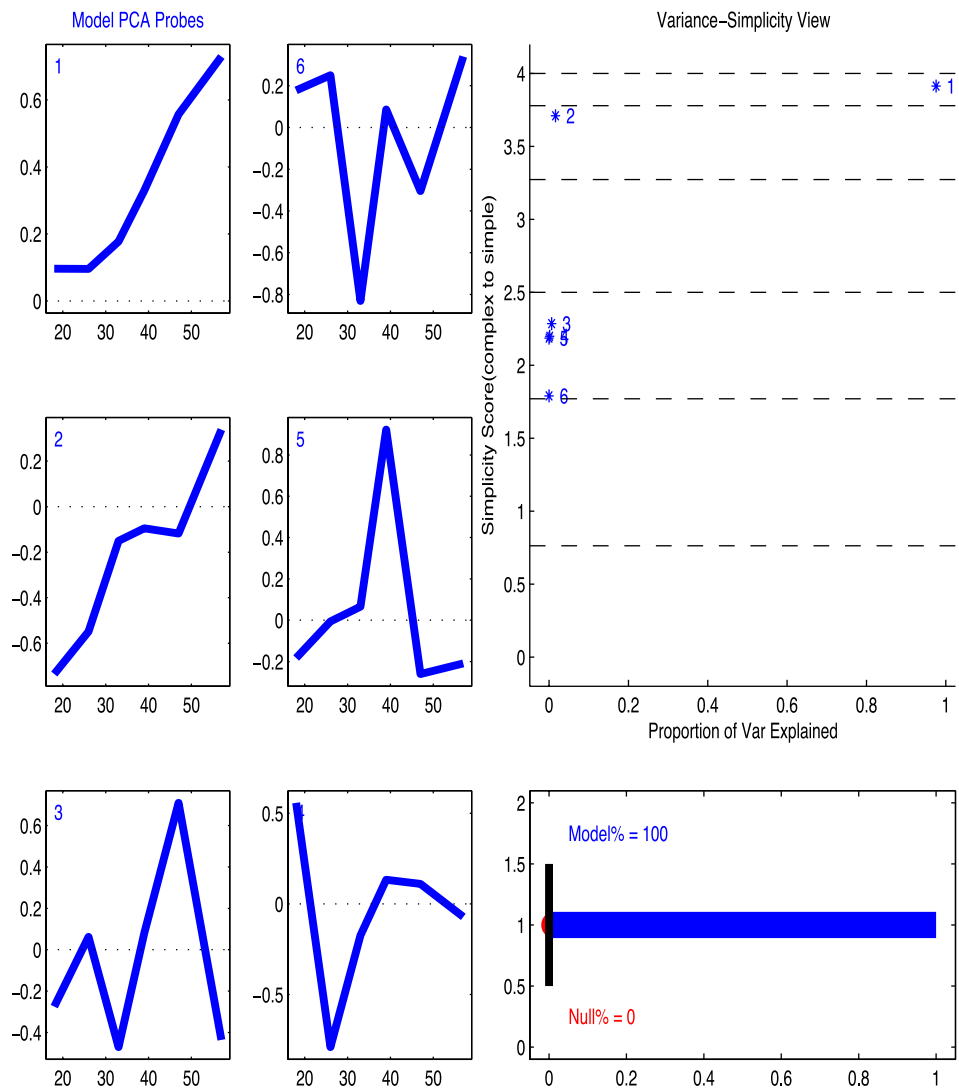}

\caption{Jewelweed data for height as a function of time in the
sun/high density group: PCA basis. The structure of the plot is as in Figure
\protect\ref{figcatpca}. The amounts of genetic variance explained by
vectors 1 through 6 are, respectively, 48.98, 0.82, 0.33, 0.08, 0 and 0.}
\label{figjewelpca}
\end{figure}

\begin{figure}

\includegraphics{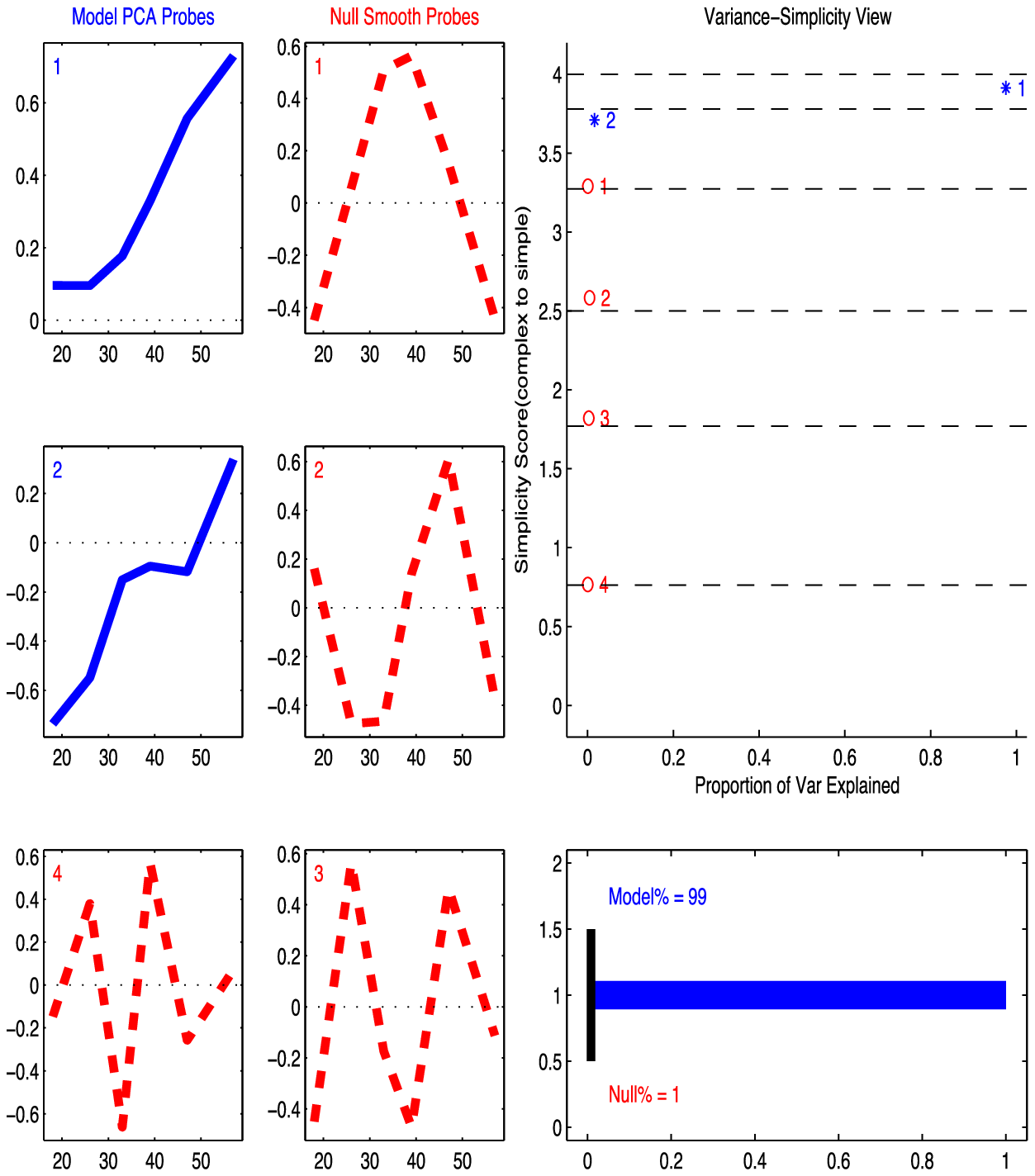}

\caption{Jewelweed data for height as a function of time in the
sun/high density group: four-dimensional nearly null space. The
structure of the plot is the same as that of Figure
\protect\ref{figcatsimplicity2}. The amounts of genetic variance explained
by simplicity vectors 1 through 4 are, respectively, 0.02, 0.22,
0.15 and 0.02.} \label{figjewelsimplicity4}
\end{figure}

The details and interpretations of Figures \ref{figcatpca} through \ref
{figjewelsimplicity4}
are in Sections
\ref{sectioncaterpillar} and~\ref{sectionjewelweed}, but we provide an
overview here.
Figures \ref{figcatpca} and \ref{figjewelpca} show the principal
component vectors for the two data
sets, corresponding to choosing $J=6$, for a six-dimensional model
space and a zero-dimensional nearly null space.
These figures are shown so we can contrast insight from a usual PC
analysis with the insight obtained from Figures
\ref{figcatsimplicity2} and \ref{figjewelsimplicity4}.
Figure \ref{figcatsimplicity2} shows the four-dimensional model space
and two-dimensional nearly null space
for the caterpillar data.
Figure \ref{figjewelsimplicity4} shows the two-dimensional model space
and four-dimensional
nearly null space for the jewelweed data.

%f4 #&#
%

%f5 #&#
%

%f6 #&#
%

The six plots in the left sides of Figures \ref{figcatpca} through \ref
{figjewelsimplicity4} show six orthonormal basis vectors for $\Re^6$.
The first $J$, in solid blue lines, are the first $J$ principal
component vectors, labeled with a blue 1 for the first principal
component, a blue 2 for the second principal component, etc. The
remaining (6--$J$) basis vectors---the dashed red lines which only
appear in Figures \ref{figcatsimplicity2} and
\ref{figjewelsimplicity4}---form the simplicity basis for the nearly
null space. The simplest basis vector is labeled with a red 1, the next
simplest with a red 2, etc. The simplicity measure is that in
(\ref{eqfirstdivided}), with large values of the measure being most
simple. We have arranged the plots of the six basis vectors so that the
top row contains what are arguably the most interesting basis vectors:
the eigenvector corresponding to the largest eigenvector and the nearly
null space's simplest vector. Nearly null space vectors are plotted
counter-clockwise in order of simplicity, while eigenvectors are
plotted clockwise in order of the corresponding eigenvalues.

Each of the six basis vectors, when used as a selection gradient,
produces an expected response to selection, as given in equation
(\ref{Breedersbeta}).
A natural estimate of the expected response to selection
for a selection gradient $\beta$ is simply $\hat{\mathcal{G}}
\beta$, where $\hat{\mathcal{G}}$ is the estimate of the genetic
covariance matrix. The captions under Figures \ref{figcatpca} through
\ref{figjewelsimplicity4}
list the six vector norms of $\hat{\mathcal{G}}\beta$, one for each of
the six basis vectors.
Note that, if $\beta$ is a unit-length eigenvector of $\hat{\mathcal
{G}}$, then the norm of $\hat{\mathcal{G}}
\beta$ is simply equal to the associated eigenvalue. The norm of $\hat
{\mathcal{G}} \beta$ is maximal when $\beta$ is the eigenvector associated
with the largest eigenvalue of $\hat{\mathcal{G}}$.

The norms of the expected responses to selection can also be reported
as proportions of genetic variance, simply by reporting each norm
divided by
the sum of the norms.
These proportions of genetic variance
are plotted on the horizontal axes in the upper right panels of Figures
\ref{figcatpca} through \ref{figjewelsimplicity4},
with simplicity scores on the vertical axes.
The numbering and color-coding correspond to the plots on the left.
The panel in
the lower right shows the total genetic variance explained by the model
space and by the
nearly null space.\looseness=-1

%s5.1 #&#
\subsection{Caterpillar data analysis}
\label{sectioncaterpillar}

Kingsolver, Ragland and Shlichta (\citeyear{KRS2004}) estimated genetic
variation in short-term growth rates of caterpillars at several
temperatures ranging from 11$^{\circ}$C to 40$^{\circ}$C. The type of
caterpillar was \textit{Pieris rapae}, which develops into the Small
Cabbage White Butterfly. The caterpillars cause extensive damage to
crops such as cabbage and broccoli, so understanding their growth is
important for commercial agriculture. The goal of the study was to
quantify patterns of genetic variation in growth rate across
temperatures, and explore how these patterns might affect evolutionary
responses to selection in changing temperature conditions in nature.
For instance, if there was little genetic variability in growth rates
at high temperatures, rising temperatures could cause extinction of
Small Cabbage White Butterflies. Alternatively, if growth rate at high
temperatures is negatively genetically correlated with growth rate at
low temperatures, then rising temperatures could lead to reductions in
growth rate at low temperatures.\looseness=1

Caterpillars were reared individually from hatching on artificial diet
in diurnally fluctuating conditions of
temperature (11--35$^{\circ}$C) and light (15 hours of light,
9 hours of dark) until the start of the developmental stage known as
the 4th larval instar. The studies focused on the 4th instar because
more than 85\% of all growth occurs during the 4th and 5th instars;
measurements were concentrated within a single instar to quantify
effects of temperature on larval growth (mass increase) as distinct
from development (molting, or the developmental processes involved in
the transitions between instars). See \citet{KRS2004} for details
of the measurements and methods. Briefly, the short-term growth rate of
each caterpillar was measured at six different temperatures between
11$^{\circ}$C and 40$^{\circ}$C (11, 17, 23, 29, 35 and 40$^{\circ}$C),
during the first two days of the 4th instar (Figure \ref{Catdata}). To
reflect the natural diurnal cycle typically experienced by caterpillars
in nature, measurements at higher temperatures were done during the day
(light phase) and at lower temperatures during the night (dark phase);
growth rate was calculated as the net change in mass over the
measurement period. Because exposure to 40$^{\circ}$C is potentially
stressful and could affect subsequent feeding and growth, measurements
at this temperature were done last for each caterpillar. Measurements
of 1088 individuals from 90 independent families of full siblings were
completed. These data were used to estimate the genetic
variance--covariance matrix ${\mathcal{G}}$ for growth rate at the six
temperatures, using the REML software \textit{dfreml} described by
\citet{MeyerSmith}. %and \citet{Meyer1998}.

To study the nearly null space, we define simplicity via (\ref
{eqfirstdivided}) with $t_j-t_{j-1}=6$
for $j=2,\ldots,5$ and $t_6-t_5=5$.

Figure \ref{figcatpca} shows the principal components decomposition
(six-dimensional model space, 0-dimensional null space) of the matrix
${\mathcal{G}}$. The first PC, explaining 59.5\% of the variation, is
dominated by strong loadings of opposite sign for growth rate at
35$^{\circ}$C and 40$^{\circ}$C, reflecting the strong negative genetic
covariance between growth at these two temperatures. This first PC has
a low simplicity score. In contrast, the second PC has a much higher
simplicity score and reflects loadings of the same sign and similar
magnitude for growth across most temperatures (17--40$^{\circ}$C). Note
that the first three PCs, totaling over 93\% of the variation, have
small loadings for growth rate at 11$^{\circ}$C, reflecting the low
genetic variation at the lowest temperature.

Figures 1 to 7 in the supplementary material [\citet{Suppplots}]
illustrate results for these data for model and null spaces of
different dimensionality (from 0 to 6 dimensions). For purposes of
discussion we focus on results for the four-dimensional model space and
two-dimensional null space (Figure \ref{figcatsimplicity2}): here the
null space includes less than 1\% (0.7\%) of the total genetic
variance, sufficiently small to strongly constrain rates of
evolutionary responses. The simplest vector in the null space is a
contrast between large loadings at lower temperatures (11--23$^{\circ
}$C) and smaller loadings of opposite sign at higher temperatures
(35--40$^{\circ}$C). We can interpret this direction in the genetic null
space in terms of lack of evolutionary response to selection:
simultaneous selection for increased growth rate at lower temperatures
and for decreased growth rate at high temperatures would result in very
little evolutionary change, because of the lack of genetic variation in
this direction.

It is also informative to consider the simplicity decomposition of the
${\mathcal{G}}$-matrix, in this case when the null-space is
six-dimensional (Supplementary Figure 7). For example, about 18\% of the
variance is associated with the simplest possible direction, for which
loadings are equal across all temperatures. This direction represents
variation in overall growth rate independent of temperature [\citet
{KGC2001,Izem2005}]. Because overall growth rate may be positively
related to fitness in a variety of situations, selection in this
direction may occur frequently in nature; the simplicity analysis
quantifies the genetic variation and the predicted evolutionary
response to such selection.

%s5.2 #&#
\subsection{Jewelweed data analysis}
\label{sectionjewelweed}

In a study of the genetic variability of height in different environments,
\citet{Stinchcombe2010} measured the heights of individuals of the
North American annual plant \textit{Impatiens capensis}
(jewelweed) in ten different greenhouse environments gotten from all
combinations of two light treatments (sun and shade) and five density
environments ranging from 64 plants per square meter to 1225 plants per
square meter.
Individuals' heights were measured to the nearest millimeter at six
time points: 18, 26, 33, 39, 47 and 57 days.

Here we analyze just one portion of the data: height as a function of
time for plants grown in sun at density 1225 plants per square meter.
The analysis appears in more detail in \citet{Gaydosthesis}.
Our purpose is to study the genetic variability in growth curves in
this environment.
Genetic variability will allow the plants to adapt to a range of
conditions, such as sunlight (taller
than average plants are typically favored) or the presence of high winds.

The estimate of $\mathcal{G}$, the genetic covariance matrix, was
produced using SAS PROC MIXED. Although the estimate is called a REML
estimate, no restrictions were placed on the estimate to ensure it
would be nonnegative definite. The resulting estimate of $\mathcal{G}$
had two eigenvalues that were negative but close to 0 (values of $-$0.21
and $-$0.55), very small compared to the value of
the largest eigenvalue (value of 183.7). We set the two negative
eigenvalues equal to 0 and calculated our final estimate
of $\mathcal{G}$, using the eigen-expansion based on the remaining
four eigenvalues and eigenvectors. Figures 8 through 14 in the supplementary
material [\citet{Suppplots}] illustrate our results for null
spaces of dimension 0 through~6.

The first principal component explains 97.5\% of the
variance and the first and second principal components explain 99.2\%
of the variance.
Based on the interpretability of the first two PCs and the proportion
of variance explained, we recommend
using a two-dimensional model space and four-dimensional nearly null
space, displayed in Figure \ref{figjewelsimplicity4}.
To find the simplicity basis of the nearly null space, we define
simplicity via (\ref{eqfirstdivided})
with the $t_j-t_{j-1}$'s equal to 8, 7, 6, 8 and 10, the differences
in the time points.

The first principal component (see Figure \ref{figjewelpca}) has a very
small loading on early ages
with loadings increasing as the plant ages. Thus, using just
the first principal component, we see that,
in a sunny dense environment, the population will be able to
evolve and
adapt to a wide range of forces of selection that act on late-age
heights. Such genetic
variation would be important if late season height is under natural
selection---for example,
if plants that are larger late in the season are able to acquire more
light and more successfully mature their seeds.

The second principal component indicates that there is some genetic
variability at young ages.

The simplest direction in the nearly null space, labeled with a red 1
in Figure \ref{figjewelsimplicity4},
shows that there is little genetic variation in the contrast of
late/early life heights to mid-life heights.
With this lack of genetic variation, the species will not be able to
adapt when the variability of environmental conditions
is in the form of a contrast between early/late season and mid-season.
For instance, if a typical season begins and ends with little sunshine,
but has high winds in mid-season, selection might favor
plants that are taller than average at the beginning and end of the
season, but shorter than average mid-season; that this combination of
traits is in the null space, however, suggests that there would be
little to no evolutionary response to such
seasonal conditions.

Note the additional insight gained by considering the simplicity basis
over simply considering
PC analysis, shown in Figure \ref{figjewelpca}. Interpreting PCs 3
through 6 is much harder than interpreting the simplest element of
their span,
that is, the simplest element of the nearly null space. While one might
infer from PCs 1 and 2 that the simplest
vector in the nearly null space is close to a parabola, our more
rigorous approach confirms that ad hoc insight.
In addition, the graphical plot in Figure~\ref{figjewelsimplicity4}
gives equal importance to the structure of the
first few PCs and the structure of the nearly null space.

%s6 #&#
\section{Simulation study}
\label{sectsimulation}

We carried out a simulation study to get insight about the effects of
sampling variation, in particular, how it affects the shape of the
simplest vector and predictions about selection response.

To reduce computational burden, our design and our estimate of the
genetic covariance were very simple. We used a balanced design with
$N_f=100$ independent
families and $n=20$ half-siblings within each family. We estimated
the genetic covariance matrix via the classic ANOVA/method of moments.
See Chapter 18 of \citet{WalshLynch}. This method leads to a
closed form estimate of the genetic covariance matrix, but
can only be used in simple designs.

We generated data for individual $i$ of family $j$ according to $y_{ij}
= \mu+ g_{ij} + e_{ij} + \varepsilon_{ij}$,
where $g_{ij}$, $e_{ij}$ and $\varepsilon_{ij}$ were independent normal
vectors of length $K=6$, with means equal to the zero vector
and with covariance matrices denoted ${\cal{G}}$, ${\cal{E}}$ and
$\sigma^2$I, respectively. We set all parameters equal to the estimates
from the caterpillar study of \citet{KRS2004},
as described in Section \ref{sectioncaterpillar}.

We simulated 200 data sets and studied three-dimensional nearly null
spaces. Each simulated data set yielded an estimated
genetic covariance matrix with its eigenvalues and eigenvectors, an
estimated nearly null space, the simplest vector in that estimated
nearly null space and the expected response to selection
under a selection gradient equal to the simplest vector.

The ANOVA method can lead to a negative definite genetic covariance
matrix estimate.
For our estimated 6 by 6 genetic covariance matrices, all 200 had the
first five eigenvalues positive but for 170 of the 200,
the smallest eigenvalue was negative. We adjusted these 170 estimates
of ${\cal{G}}$ by setting the 170 smallest eigenvalues to 0. As the
magnitudes of the negative eigenvalues were small (the smallest
eigenvalue was $-$0.068), this adjustment had little impact. Note that
resetting the eigenvalues leaves the eigenvectors unchanged.

\begin{figure}

\includegraphics{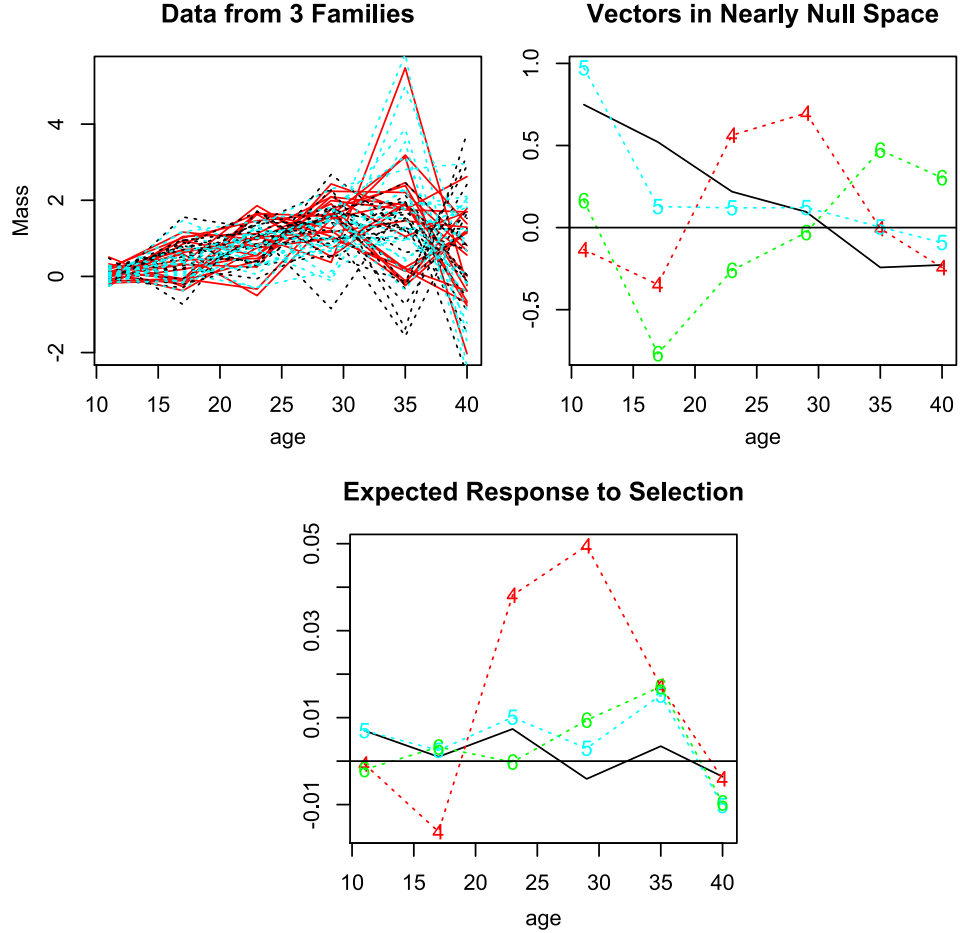}
 \caption{Results from the first simulated data set.
The upper left plot shows data from three of the 100 families. The
upper right plot shows vectors in the estimated nearly null space (the
simplest vector in black and three PCs---PC4, PC5 and PC6). The lower
left plot shows the expected responses to selection when the selection
gradient is equal to each of the vectors in the upper right plot.}
\label{figdatafamily}
\end{figure}

Figure \ref{figdatafamily} provides information from one of the 200
simulated data sets.
The upper left plot shows the simulated data from three of the one
hundred families, color-coded by family. The upper right plot shows
four vectors in the estimated nearly null space: the
black line is the simplest vector and the remaining lines are the
fourth, fifth and sixth principal components of the estimated genetic
covariance matrix.
The lower left plot shows the expected response to selection when using
the four depicted vectors in the nearly null space as selection gradients.
The expected response to selection is calculated using
the ``true genetic covariance matrix,'' that is,
the genetic covariance matrix used to generate the simulated data.
The magnitudes of the vectors of expected responses to selection are 0.012
for the simplest vector and
0.067, 0.022 and 0.021 for the three principal components. The largest
possible magnitude of the expected response to selection is
the largest eigenvalue of the true genetic covariance matrix, that is, 0.618.

%f7 #&#
%

\begin{figure}

\includegraphics{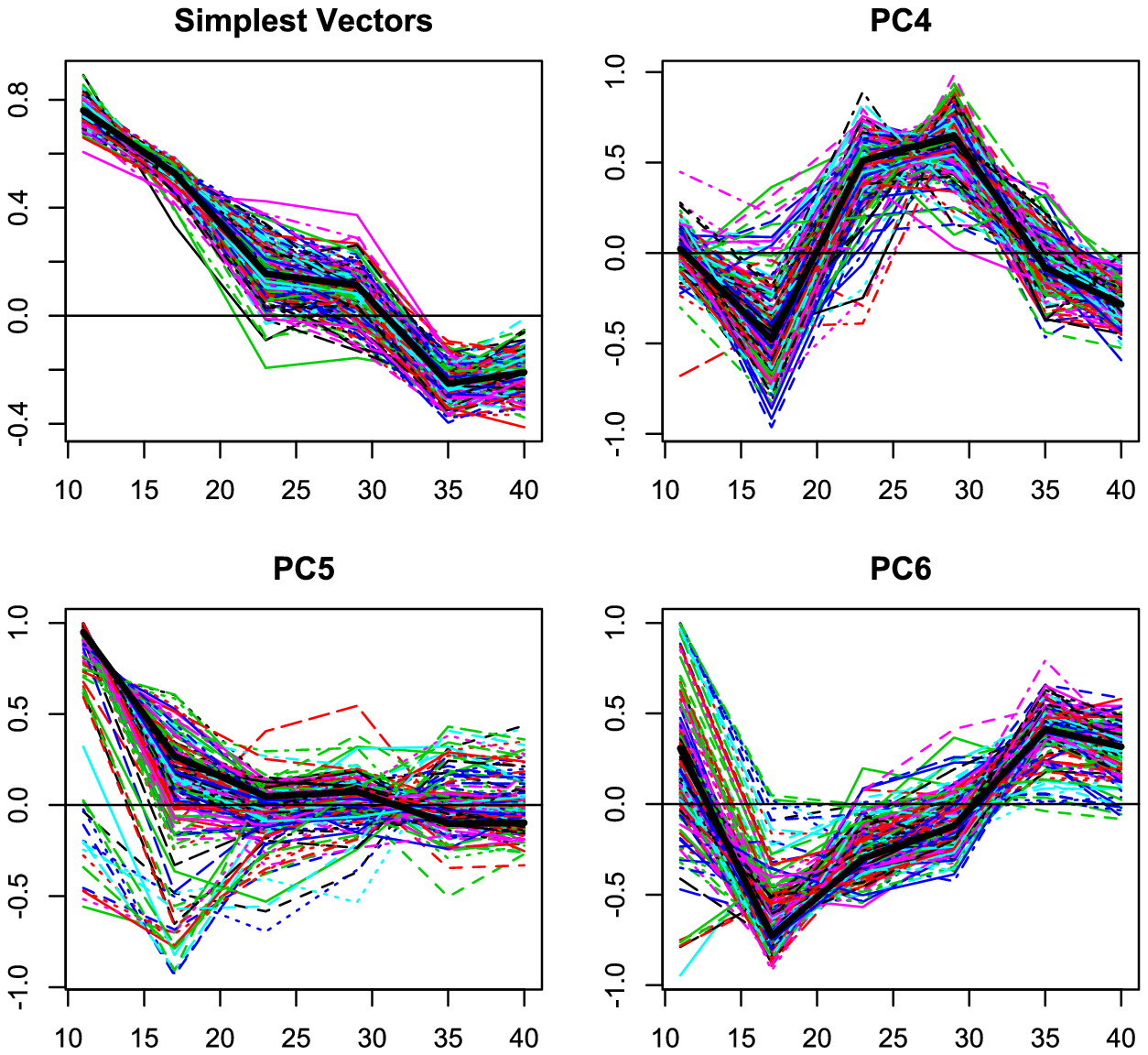}

\caption{Simulation results from 200 data sets: the 200 simplest
vectors in the estimated three-dimensional nearly null space and
the eigenvectors corresponding to the three smallest eigenvalues of
the estimated genetic covariance matrix along with the true
eigenvectors (remaining three plots). In each plot, a true vector
appears as a dark thick line. Recall that
an eigenvector or simplicity vector is only defined up to a multiple
of $+/{-}1$. Multipliers have been chosen so that the vectors in the above
plots are similar.}
\label{figvectors}
\end{figure}

\begin{figure}

\includegraphics{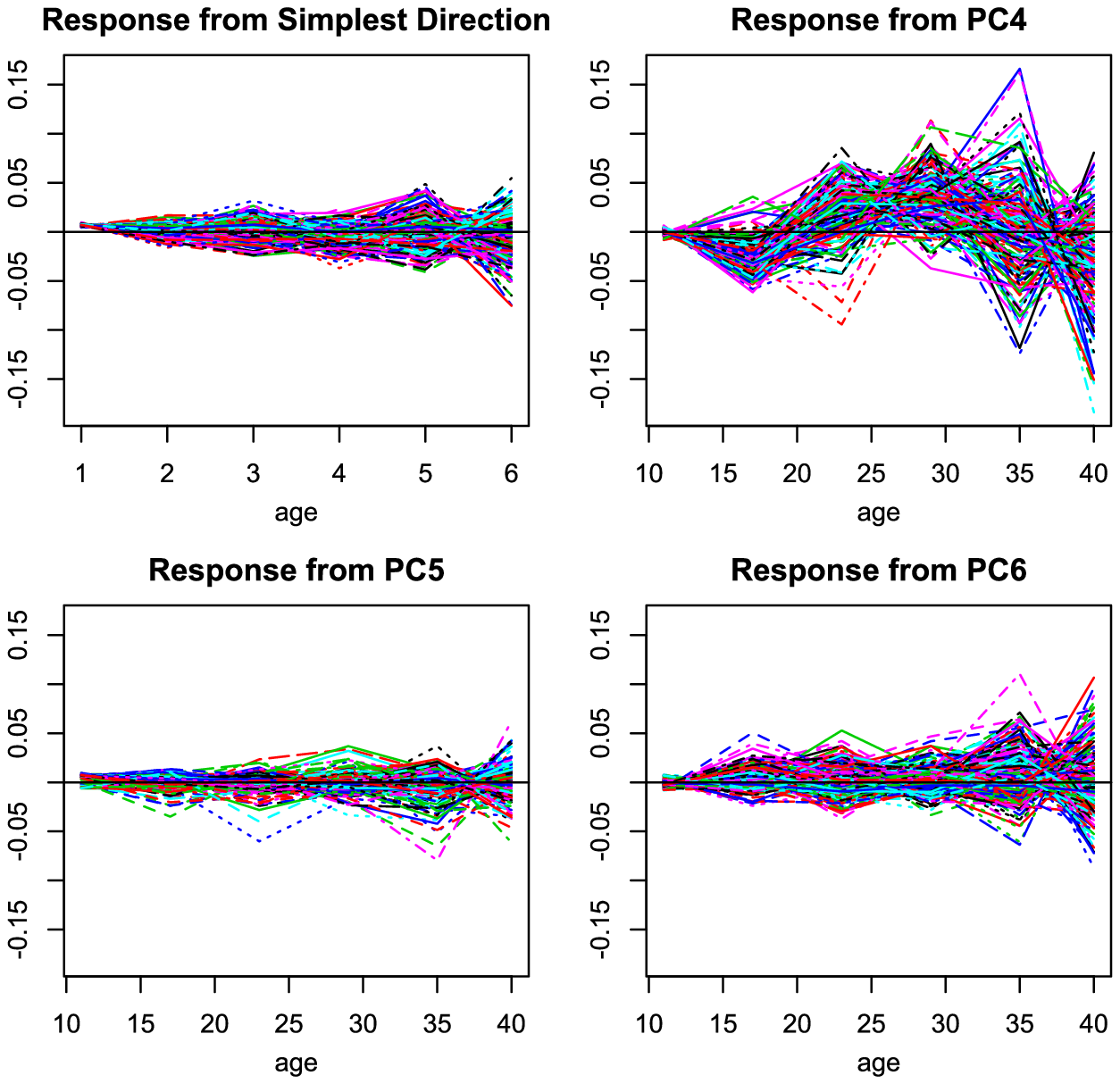}

\caption{Simulation results from 200 data sets: expected response to
selection from the simplest vector in the estimated nearly null space
and the three PCs in the estimated nearly null space.}
\label{figresponses}
\end{figure}

Figures \ref{figvectors} and \ref{figresponses} contain the results of
our simulation study. In Figure \ref{figvectors} the upper left plot
shows the 200 simplest vectors in the estimated nearly null space. The
other three plots in that figure show the eigenvectors that span the
estimated nearly null space. The
upper right plot contains the 200 ``fourth eigenvectors,'' that is,
those corresponding to the fourth largest eigenvalues of the estimated
genetic covariance matrices. The lower left plot contains the 200 ``fifth
eigenvectors'' and the lower right plot contains the 200 ``sixth
eigenvectors.''

%f8 #&#
%

%f9 #&#
%

In Figure \ref{figresponses}, the upper right plot contains the
200 expected responses to selection calculated using the 200 simplest
vectors of Figure \ref{figvectors} as selection gradients and the
``true'' genetic covariance matrix.
The remaining plots contain
the expected responses to selection calculated using the eigenvectors
shown in Figure \ref{figvectors} as selection gradients.

From Figures \ref{figdatafamily} to \ref{figresponses}, we can see that
the simplest vectors in the estimated nearly null spaces are always
interpretable and send the same clear
message. In contrast, the fourth principal components (the
``dominant'' component in the estimated nearly null spaces) are
difficult to interpret, as we expected.
The simplest vectors vary little from data set to data set and, when
used as
selection gradients, the simplest vectors yield expected responses to
selection that are close to 0, with little variability (mean
length of the response vectors is 0.032 with standard deviation 0.001).
In contrast, when the fourth eigenvector is the selection gradient,
the magnitudes of the expected response vectors
are larger and more variable, with mean length 0.091 and standard
deviation 0.007.

%s7 #&#
\section{Theory}

The asymptotic consistency of our method follows directly from consistency
of the estimated genetic covariance matrix.
Under conditions, the REML estimate $\hat{\mathcal{G}}$ is
asymptotically equivalent to the maximum likelihood estimate, which converges
in probability to $\mathcal{G}$, the true genetic covariance matrix
[\citet{Demidenko}, page 181].
In this case, the eigenvalues of $\hat{\mathcal{G}}$ converge to those
of $\mathcal{G}$,
since eigenvalues
are defined as solutions of the characteristic polynomial, and
the coefficients of the characteristic polynomial of $\hat{\mathcal
{G}}$ converge to those of $\mathcal{G}$.
Thus, for many common methods of estimating the dimension $J$ of
the model space, the dimension of
the estimated model space converges to $J$, with, possibly,
the requirement that $\lambda_J > \lambda_{J+1}$.
For instance,
the convergence of estimated eigenvalues implies
that the proportion of variance explained by the first $J$ eigenvectors
of $\hat{\mathcal{G}}$ will converge to the proportion of variance
explained by
the first $J$ eigenvectors of $\mathcal{G}$.\looseness=-1

Showing convergence of estimated eigenspaces requires more care due to
the complication of
defining distances between subspaces and due to the possibility of
multiplicity of the roots of the characteristic polynomial and the resulting
nonuniqueness of eigenvectors. See
\citet{Gaydosthesis}, who shows that, under conditions, the
nearly null space
of the usual sample covariance matrix converges to that of the true
covariance matrix. To define
convergence, Gaydos
defines the squared distance between two subspaces as
the sum of the squared sines of
the canonical angles between the two subspaces.
See \citet{StewartSun1990} for a discussion of canonical angles.

%s8 #&#
\section{Discussion}

We have proposed simplicity measures and developed accompanying
graphical tools to explore and visualize directions of low variability
in vector-valued traits. The techniques allow us to more directly study
the space spanned by the lowest variance
PCs. When examined individually, these PCs typically have little
structure. Considering them jointly as a subspace
allows us to find the simplest structure within that subspace. Our
graphical tools allow us to consider subspaces of
different dimensions, easily seeing the simplicity and variance
explained by the subspace and individual vectors.

Here, we have studied the nearly null space by defining a simplicity
basis with the simplicity
of a vector $v$ of the form
$v' \Lambda v$, and we have analyzed data with the simplicity of $v$
defined in terms
of first divided differences of the components of $v$.
Instead of using such a smoothing-based simplicity measure, one could
consider a sparseness measure, deeming a
vector to be simple if it has many zero components.
In a modification of principal components, \citet{Chipman2005}
define sparseness of a vector in terms
of the number of its nonzero
components.
Another sparseness measure, used in the varimax method of factor
rotation in factor analysis
[\citet{JohnsonandWichern}], defines a quadratic measure of
sparseness of the vector $v$, namely,
$\sum(\vecv_i - \bar{\vecv})^2$, with large values indicating greater
simplicity.
An $L_1$ measure of sparseness, namely, $\sum|\vecv_i|$, is used in
the Lasso
technique for
regression
[\citet{Tibshirani1996}], with small values indicating greater simplicity.

Our methodology can,
in principle, be extended
to function-valued traits.
Genetic constraints can be defined for function-valued traits via the
work of \citet{KirkpatrickandHeckman},
\citet{GomulkiewiczandBeder} and \citet{BederandGomulkiewicz},
who showed the
validity of the Breeder's equation in (\ref{Breeders}) and (\ref{Breedersbeta})
when the phenotype is a function.
The advantages of functional data analysis
techniques over multivariate techniques are well known in the
statistical literature. For instance,
functional data analysis
does not require that individuals be measured at the same time points
or even at the same number of times.
Furthermore, functional data analysis uses the smoothness underlying
the data to avoid high-dimensional analysis
problems caused by a
large number of observations per individual.
The advantages of functional data analysis are only
just catching hold in the biological literature. See, for instance,
\citet{GomulkiewiczandKingsolver} and \citet{Griswold2008}.

Defining a simplicity basis for the nearly null space is especially
useful in the analysis of functional data. To see this, suppose that
the genetic component $g$
is a continuous time random process. Then, under conditions, we can
write $g$ in terms
of its Karhunen--Lo\'eve expansion: $g(t) = \sum_1^{\infty} \alpha_j
\phi
_j(t)$, where the
$\phi_j$'s are orthonormal functions and the $\alpha_j$'s are
independent with
mean zero and variances $\lambda_1 \ge\lambda_2 \ge\cdots\,$. See, for instance,
\citet{LoeveII} or \citet{Adler2007}.
If our model for $g$ allows a countably infinite number of these
variances to be positive, then the true model space for $g$ is infinite
dimensional.
However, since any particular data set is finite dimensional, estimates
of $g$ always lie in a
finite-dimensional space. Hence, the estimated model space is
finite dimensional and its orthogonal complement
is infinite dimensional. A natural way to study
this infinite-dimensional subspace is by finding its simplest
directions and seeing if these
directions have any interpretable structure.

Studying the structure of low variance subspaces can provide biologists
with insights into the existence of genetic constraints.
But the notion of a simplicity basis and the associated visualization
tools may be useful
in other contexts, in particular, in providing modeling tools in the
analysis of smooth high-dimensional data.

\begin{supplement}%[id=suppA]
\sname{Supplement A}
\stitle{Supplementary plots\\}
\slink[doi]{10.1214/12-AOAS603SUPPA} %[doi,text={...}] - jei reikia
%suskaldyti doi
\sdatatype{.pdf}
\sfilename{aoas603\_suppa.pdf}
\sdescription{As previously noted, supplementary material
[\citet{Suppplots}] contains a complete set of plots from our data
analyses, as in Figures \ref{figcatpca} through \ref{figjewelsimplicity4}.}
\end{supplement}

\begin{supplement}%[id=suppA]
\sname{Supplement B}
\stitle{Nearly null space example\\}
\slink[doi]{10.1214/12-AOAS603SUPPB} %[doi,text={...}] - jei reikia
%suskaldyti doi
\sdatatype{.pdf}
\sfilename{aoas603\_suppb.pdf}
\sdescription{An additional supplementary file
[\citet{Suppexample}] contains a simple example that shows the
benefits of the proposed methodology.}
\end{supplement}

% imsref loaded by lrinkeviciute, 2013-01-09 12:52:07
%
% imsref loaded by lrinkeviciute, 2013-01-09 13:37:24
%

\printaddresses


\begin{thebibliography}{36}
% BibTex style file: ims.bst, 2012-08-21
% Default style options (sort=0,type=number).
% Used options (sort=1,type=nameyear).

%b1 #&#
\bibitem[\protect\citeauthoryear{Adler and Taylor}{2007}]{Adler2007}
%
\begin{bbook}[mr]
\bauthor{\bsnm{Adler},~\bfnm{Robert~J.}\binits{R.~J.}} \AND
\bauthor{\bsnm{Taylor},~\bfnm{Jonathan~E.}\binits{J.~E.}}
(\byear{2007}).
\btitle{Random Fields and Geometry}.
\bpublisher{Springer}, \blocation{New York}.
\bid{mr={2319516}}
\bptok{imsref}%
\end{bbook}
%
\endbibitem

%b2 #&#
\bibitem[\protect\citeauthoryear{Amemiya, Anderson and
Lewis}{1990}]{Amemiya1990}
%
\begin{barticle}[mr]
\bauthor{\bsnm{Amemiya},~\bfnm{Yasuo}\binits{Y.}},
\bauthor{\bsnm{Anderson},~\bfnm{T.~W.}\binits{T.~W.}} \AND
\bauthor{\bsnm{Lewis},~\bfnm{Peter A.~W.}\binits{P.~A.~W.}}
(\byear{1990}).
\btitle{Percentage points for a test of rank in multivariate components of
variance}.
\bjournal{Biometrika}
\bvolume{77}
\bpages{637--641}.
\bid{doi={10.1093/biomet/77.3.637}, issn={0006-3444}, mr={1087855}}
\bptok{imsref}%
\end{barticle}
%
\endbibitem

%b3 #&#
\bibitem[\protect\citeauthoryear{Anderson and
Amemiya}{1991}]{AndersonandAmemiya}
%
\begin{barticle}[mr]
\bauthor{\bsnm{Anderson},~\bfnm{T.~W.}\binits{T.~W.}} \AND
\bauthor{\bsnm{Amemiya},~\bfnm{Yasuo}\binits{Y.}}
(\byear{1991}).
\btitle{Testing dimensionality in the multivariate analysis of variance}.
\bjournal{Statist. Probab. Lett.}
\bvolume{12}
\bpages{445--463}.
\bid{doi={10.1016/0167-7152(91)90002-9}, issn={0167-7152}, mr={1143744}}
\bptok{imsref}%
\end{barticle}
%
\endbibitem

%b4 #&#
\bibitem[\protect\citeauthoryear{Beder and
Gomulkiewicz}{1998}]{BederandGomulkiewicz}
%
\begin{barticle}[mr]
\bauthor{\bsnm{Beder},~\bfnm{Jay~H.}\binits{J.~H.}} \AND
\bauthor{\bsnm{Gomulkiewicz},~\bfnm{Richard}\binits{R.}}
(\byear{1998}).
\btitle{Computing the selection gradient and evolutionary response of an
infinite-dimensional trait}.
\bjournal{J. Math. Biol.}
\bvolume{36}
\bpages{299--319}.
\bid{doi={10.1007/s002850050102}, issn={0303-6812}, mr={1608605}}
\bptnote{check year}%
\bptok{imsref}%
\end{barticle}
%
\endbibitem

%b5 #&#
\bibitem[\protect\citeauthoryear{Chipman and Gu}{2005}]{Chipman2005}
%
\begin{barticle}[mr]
\bauthor{\bsnm{Chipman},~\bfnm{Hugh~A.}\binits{H.~A.}} \AND
\bauthor{\bsnm{Gu},~\bfnm{Hong}\binits{H.}}
(\byear{2005}).
\btitle{Interpretable dimension reduction}.
\bjournal{J. Appl. Stat.}
\bvolume{32}
\bpages{969--987}.
\bid{doi={10.1080/02664760500168648}, issn={0266-4763}, mr={2221888}}
\bptok{imsref}%
\end{barticle}
%
\endbibitem

%b6 #&#
\bibitem[\protect\citeauthoryear{Demidenko}{2004}]{Demidenko}
%
\begin{bbook}[mr]
\bauthor{\bsnm{Demidenko},~\bfnm{Eugene}\binits{E.}}
(\byear{2004}).
\btitle{Mixed Models: Theory and Applications}.
\bpublisher{Wiley}, \blocation{Hoboken, NJ}.
\bid{doi={10.1002/0471728438}, mr={2077875}}
\bptok{imsref}%
\end{bbook}
%
\endbibitem

%b7 #&#
\bibitem[\protect\citeauthoryear{Eilers and Marx}{1996}]{EilersMarx1996}
%
\begin{barticle}[mr]
\bauthor{\bsnm{Eilers},~\bfnm{Paul H.~C.}\binits{P.~H.~C.}} \AND
\bauthor{\bsnm{Marx},~\bfnm{Brian~D.}\binits{B.~D.}}
(\byear{1996}).
\btitle{Flexible smoothing with {$B$}-splines and penalties}.
\bjournal{Statist. Sci.}
\bvolume{11}
\bpages{89--121}.
\bid{doi={10.1214/ss/1038425655}, issn={0883-4237}, mr={1435485}}
\bptnote{check related}%
\bptok{imsref}%
\end{barticle}
%
\endbibitem

%b8 #&#
\bibitem[\protect\citeauthoryear{Gaydos}{2008}]{Gaydosthesis}
%
\begin{bmisc}[author]
\bauthor{\bsnm{Gaydos},~\bfnm{Travis}\binits{T.}}
(\byear{2008}).
\bhowpublished{Data representation/basis selection to understand
variation of
function valued traits. Ph.D. thesis, Univ. North Carolina}.
\bptok{imsref}%
\end{bmisc}
%
\endbibitem

%b9 #&#
\bibitem[\protect\citeauthoryear{Gaydos et~al.}{2013a}]{Suppplots}
%
\begin{bmisc}[author]
\bauthor{\bsnm{Gaydos},~\bfnm{T.}\binits{T.}},
\bauthor{\bsnm{Heckman},~\bfnm{N.}\binits{N.}},
\bauthor{\bsnm{Kirkpatrick},~\bfnm{M.}\binits{M.}},
\bauthor{\bsnm{Stinchcombe},~\bfnm{J.~R.}\binits{J.~R.}},
\bauthor{\bsnm{Schmitt},~\bfnm{J.}\binits{J.}},
\bauthor{\bsnm{Kingsolver},~\bfnm{J.}\binits{J.}} \AND
\bauthor{\bsnm{Marron},~\bfnm{J.~S.}\binits{J.~S.}}
(\byear{2013}a).
\bhowpublished{Supplement to ``Visualizing genetic constraints.''
DOI:\doiurl{10.1214/12-AOAS603SUPPA}}.
\bptok{imsref}%
\end{bmisc}
%
\endbibitem

%b10 #&#
\bibitem[\protect\citeauthoryear{Gaydos et~al.}{2013b}]{Suppexample}
%
\begin{bmisc}[author]
\bauthor{\bsnm{Gaydos},~\bfnm{T.}\binits{T.}},
\bauthor{\bsnm{Heckman},~\bfnm{N.}\binits{N.}},
\bauthor{\bsnm{Kirkpatrick},~\bfnm{M.}\binits{M.}},
\bauthor{\bsnm{Stinchcombe},~\bfnm{J.~R.}\binits{J.~R.}},
\bauthor{\bsnm{Schmitt},~\bfnm{J.}\binits{J.}},
\bauthor{\bsnm{Kingsolver},~\bfnm{J.}\binits{J.}} \AND
\bauthor{\bsnm{Marron},~\bfnm{J.~S.}\binits{J.~S.}}
(\byear{2013}b).
\bhowpublished{Supplement to ``Visualizing genetic constraints.''
DOI:\doiurl{10.1214/12-AOAS603SUPPB}}.
\bptok{imsref}%
\end{bmisc}
%
\endbibitem

%b11 #&#
\bibitem[\protect\citeauthoryear{Gomulkiewicz and
Beder}{1996}]{GomulkiewiczandBeder}
%
\begin{barticle}[mr]
\bauthor{\bsnm{Gomulkiewicz},~\bfnm{Richard}\binits{R.}} \AND
\bauthor{\bsnm{Beder},~\bfnm{Jay~H.}\binits{J.~H.}}
(\byear{1996}).
\btitle{The selection gradient of an infinite-dimensional trait}.
\bjournal{SIAM J. Appl. Math.}
\bvolume{56}
\bpages{509--523}.
\bid{doi={10.1137/S0036139993255765}, issn={0036-1399}, mr={1381657}}
\bptok{imsref}%
\end{barticle}
%
\endbibitem

%b12 #&#
\bibitem[\protect\citeauthoryear{Gomulkiewicz and
Houle}{2009}]{GomulkiewiczandHoule}
%
\begin{barticle}[author]
\bauthor{\bsnm{Gomulkiewicz},~\bfnm{R.}\binits{R.}} \AND
\bauthor{\bsnm{Houle},~\bfnm{D.}\binits{D.}}
(\byear{2009}).
\btitle{Demographic and genetic constraints on evolution}.
\bjournal{American Naturalist}
\bvolume{174}
\bpages{218--229}.
\bptok{imsref}%
\end{barticle}
%
\endbibitem

%b13 #&#
\bibitem[\protect\citeauthoryear{Gomulkiewicz and
Kingsolver}{2006}]{GomulkiewiczandKingsolver}
%
\begin{barticle}[author]
\bauthor{\bsnm{Gomulkiewicz},~\bfnm{R.}\binits{R.}} \AND
\bauthor{\bsnm{Kingsolver},~\bfnm{J.~G.}\binits{J.~G.}}
(\byear{2006}).
\btitle{A fable of four functions: Function-valued approaches in evolutionary
biology}.
\bjournal{Journal of Evolutionary Biology}
\bvolume{20}
\bpages{20--21}.
\bptok{imsref}%
\end{barticle}
%
\endbibitem

%b14 #&#
\bibitem[\protect\citeauthoryear{Green and
Silverman}{1994}]{SilvermanGreen1994}
%
\begin{bbook}[mr]
\bauthor{\bsnm{Green},~\bfnm{P.~J.}\binits{P.~J.}} \AND
\bauthor{\bsnm{Silverman},~\bfnm{B.~W.}\binits{B.~W.}}
(\byear{1994}).
\btitle{Nonparametric Regression and Generalized Linear Models: A Roughness
Penalty Approach}.
\bseries{Monographs on Statistics and Applied Probability}
\bvolume{58}.
\bpublisher{Chapman \& Hall}, \blocation{London}.
\bid{mr={1270012}}
\bptok{imsref}%
\end{bbook}
%
\endbibitem

%b15 #&#
\bibitem[\protect\citeauthoryear{Griswold, Gomulkiewicz and
Heckman}{2008}]{Griswold2008}
%
\begin{barticle}[pbm]
\bauthor{\bsnm{Griswold},~\bfnm{Cortland~K.}\binits{C.~K.}},
\bauthor{\bsnm{Gomulkiewicz},~\bfnm{Richard}\binits{R.}} \AND
\bauthor{\bsnm{Heckman},~\bfnm{Nancy}\binits{N.}}
(\byear{2008}).
\btitle{Hypothesis testing in comparative and experimental studies of
function-valued traits}.
\bjournal{Evolution}
\bvolume{62}
\bpages{1229--1242}.
\bid{doi={10.1111/j.1558-5646.2008.00340.x}, issn={0014-3820}, pii={EVO340},
pmid={18266991}}
\bptok{imsref}%
\end{barticle}
%
\endbibitem

%b16 #&#
\bibitem[\protect\citeauthoryear{Heckman}{2003}]{Heckman2003}
%
\begin{bincollection}[mr]
\bauthor{\bsnm{Heckman},~\bfnm{Nancy~E.}\binits{N.~E.}}
(\byear{2003}).
\btitle{Functional data analysis in evolutionary biology}.
In \bbooktitle{Recent Advances and Trends in Nonparametric Statistics}
(\beditor{\bfnm{M.~G.}\binits{M.~G.}~\bsnm{Akritas}} \AND
\beditor{\bfnm{D.~N.}\binits{D.~N.}~\bsnm{Politis}}, eds.)
\bpages{49--60}.
\bpublisher{Elsevier}, \blocation{Amsterdam}.
\bid{doi={10.1016/B978-044451378-6/50004-1}, mr={2498232}}
\bptok{imsref}%
\end{bincollection}
%
\endbibitem

%b17 #&#
\bibitem[\protect\citeauthoryear{Hine and Blows}{2006}]{HineandBlows}
%
\begin{barticle}[pbm]
\bauthor{\bsnm{Hine},~\bfnm{Emma}\binits{E.}} \AND
\bauthor{\bsnm{Blows},~\bfnm{Mark~W.}\binits{M.~W.}}
(\byear{2006}).
\btitle{Determining the effective dimensionality of the genetic
variance--covariance matrix}.
\bjournal{Genetics}
\bvolume{173}
\bpages{1135--1144}.
\bid{doi={10.1534/genetics.105.054627}, issn={0016-6731},
pii={genetics.105.054627}, pmcid={1526539}, pmid={16547106}}
\bptok{imsref}%
\end{barticle}
%
\endbibitem

%b18 #&#
\bibitem[\protect\citeauthoryear{Izem and Kingsolver}{2005}]{Izem2005}
%
\begin{barticle}[pbm]
\bauthor{\bsnm{Izem},~\bfnm{Rima}\binits{R.}} \AND
\bauthor{\bsnm{Kingsolver},~\bfnm{Joel~G.}\binits{J.~G.}}
(\byear{2005}).
\btitle{Variation in continuous reaction norms: Quantifying directions of
biological interest}.
\bjournal{Am. Nat.}
\bvolume{166}
\bpages{277--289}.
\bid{doi={10.1086/431314}, issn={1537-5323}, pii={AN40690}, pmid={16032579}}
\bptok{imsref}%
\end{barticle}
%
\endbibitem

%b19 #&#
\bibitem[\protect\citeauthoryear{Johnson and Wichern}{2008}]{JohnsonandWichern}
%
\begin{bbook}[author]
\bauthor{\bsnm{Johnson},~\bfnm{Richard~A.}\binits{R.~A.}} \AND
\bauthor{\bsnm{Wichern},~\bfnm{Dean~W.}\binits{D.~W.}}
(\byear{2008}).
\btitle{Applied Multivariate Statistical Analysis},
\bedition{6th} ed.
\bpublisher{Pearson Education}, \blocation{Upper Saddle River}.
\bptok{imsref}%
\end{bbook}
%
\endbibitem

%b20 #&#
\bibitem[\protect\citeauthoryear{Kingsolver, Gomulkiewicz and
Carter}{2001}]{KGC2001}
%
\begin{barticle}[author]
\bauthor{\bsnm{Kingsolver},~\bfnm{Joel~G.}\binits{J.~G.}},
\bauthor{\bsnm{Gomulkiewicz},~\bfnm{Richard}\binits{R.}} \AND
\bauthor{\bsnm{Carter},~\bfnm{Patrick~A.}\binits{P.~A.}}
(\byear{2001}).
\btitle{Variation, selection and evolution of function valued traits}.
\bjournal{Genetica}
\bvolume{112--113}
\bpages{87--104}.
\bptok{imsref}%
\end{barticle}
%
\endbibitem

%b21 #&#
\bibitem[\protect\citeauthoryear{Kingsolver, Ragland and
Shlichta}{2004}]{KRS2004}
%
\begin{barticle}[pbm]
\bauthor{\bsnm{Kingsolver},~\bfnm{Joel~G.}\binits{J.~G.}},
\bauthor{\bsnm{Ragland},~\bfnm{Gregory~J.}\binits{G.~J.}} \AND
\bauthor{\bsnm{Shlichta},~\bfnm{J.~Gwen}\binits{J.~G.}}
(\byear{2004}).
\btitle{Quantitative genetics of continuous reaction norms: Thermal sensitivity
of caterpillar growth rates}.
\bjournal{Evolution}
\bvolume{58}
\bpages{1521--1529}.
\bid{issn={0014-3820}, pmid={15341154}}
\bptok{imsref}%
\end{barticle}
%
\endbibitem

%b22 #&#
\bibitem[\protect\citeauthoryear{Kirkpatrick and
Heckman}{1989}]{KirkpatrickandHeckman}
%
\begin{barticle}[mr]
\bauthor{\bsnm{Kirkpatrick},~\bfnm{Mark}\binits{M.}} \AND
\bauthor{\bsnm{Heckman},~\bfnm{Nancy}\binits{N.}}
(\byear{1989}).
\btitle{A quantitative genetic model for growth, shape, reaction norms, and
other infinite-dimensional characters}.
\bjournal{J. Math. Biol.}
\bvolume{27}
\bpages{429--450}.
\bid{doi={10.1007/BF00290638}, issn={0303-6812}, mr={1009899}}
\bptok{imsref}%
\end{barticle}
%
\endbibitem

%b23 #&#
\bibitem[\protect\citeauthoryear{Kirkpatrick and
Lofsvold}{1992}]{Kirkpatrick1992}
%
\begin{barticle}[author]
\bauthor{\bsnm{Kirkpatrick},~\bfnm{M.}\binits{M.}} \AND
\bauthor{\bsnm{Lofsvold},~\bfnm{D.}\binits{D.}}
(\byear{1992}).
\btitle{Measuring selection and constraint in the evolution of growth}.
\bjournal{Evolution}
\bvolume{46}
\bpages{954--971}.
\bptok{imsref}%
\end{barticle}
%
\endbibitem

%b24 #&#
\bibitem[\protect\citeauthoryear{Lande}{1976}]{Lande1976}
%
\begin{barticle}[author]
\bauthor{\bsnm{Lande},~\bfnm{R.}\binits{R.}}
(\byear{1976}).
\btitle{Natural selection and random genetic drift in phenotypic evolution}.
\bjournal{Evolution}
\bvolume{30}
\bpages{314--334}.
\bptok{imsref}%
\end{barticle}
%
\endbibitem

%b25 #&#
\bibitem[\protect\citeauthoryear{Lande}{1979}]{Lande1979}
%
\begin{barticle}[author]
\bauthor{\bsnm{Lande},~\bfnm{R.}\binits{R.}}
(\byear{1979}).
\btitle{Quantitative genetic analysis of multivariate evolution,
applied to
brain: Body size allometry}.
\bjournal{Evolution}
\bvolume{33}
\bpages{402--416}.
\bptok{imsref}%
\end{barticle}
%
\endbibitem

%b26 #&#
\bibitem[\protect\citeauthoryear{Lande and Arnold}{1983}]{Lande1983}
%
\begin{barticle}[author]
\bauthor{\bsnm{Lande},~\bfnm{R.}\binits{R.}} \AND
\bauthor{\bsnm{Arnold},~\bfnm{S.}\binits{S.}}
(\byear{1983}).
\btitle{The measurement of selection on correlated characters}.
\bjournal{Evolution}
\bvolume{37}
\bpages{1210--1226}.
\bptok{imsref}%
\end{barticle}
%
\endbibitem

%b27 #&#
\bibitem[\protect\citeauthoryear{Lo{\`e}ve}{1978}]{LoeveII}
%
\begin{bbook}[mr]
\bauthor{\bsnm{Lo{\`e}ve},~\bfnm{Michel}\binits{M.}}
(\byear{1978}).
\btitle{Probability Theory. {II}},
\bedition{4th} ed.
\bseries{Graduate Texts in Mathematics}
\bvolume{46}.
\bpublisher{Springer}, \blocation{New York}.
\bid{mr={0651018}}
\bptok{imsref}%
\end{bbook}
%
\endbibitem

%b28 #&#
\bibitem[\protect\citeauthoryear{Lynch and Walsh}{1998}]{WalshLynch}
%
\begin{bbook}[author]
\bauthor{\bsnm{Lynch},~\bfnm{M.}\binits{M.}} \AND
\bauthor{\bsnm{Walsh},~\bfnm{B.}\binits{B.}}
(\byear{1998}).
\btitle{Genetic Analysis of Quantitative Traits}.
\bpublisher{Sinauer}, \blocation{Sunderland, MA}.
\bptok{imsref}%
\end{bbook}
%
\endbibitem

%b29 #&#
%%
%(\byear{1998}).
%%

%b30 #&#
\bibitem[\protect\citeauthoryear{Meyer and Smith}{1996}]{MeyerSmith}
%
\begin{barticle}[author]
\bauthor{\bsnm{Meyer},~\bfnm{Karin}\binits{K.}} \AND
\bauthor{\bsnm{Smith},~\bfnm{SP}\binits{S.}}
(\byear{1996}).
\btitle{Restricted maximum likelihood estimation for animal models using
derivatives of the likelihood}.
\bjournal{Genetics Selection Evolution}
\bvolume{28}
\bpages{23--49}.
\bptok{imsref}%
\end{barticle}
%
\endbibitem

%b31 #&#
\bibitem[\protect\citeauthoryear{Schatzman}{2002}]{Schatzman2002}
%
\begin{bbook}[author]
\bauthor{\bsnm{Schatzman},~\bfnm{Michelle}\binits{M.}}
(\byear{2002}).
\btitle{Numerical Analysis: A Mathematical Introduction}.
\bpublisher{Claredon Press}, \blocation{Oxford}.
\bptok{imsref}%
\end{bbook}
%
\endbibitem

%b32 #&#
\bibitem[\protect\citeauthoryear{Searle, Casella and
McCulloch}{2006}]{Searle2006}
%
\begin{bbook}[mr]
\bauthor{\bsnm{Searle},~\bfnm{Shayle~R.}\binits{S.~R.}},
\bauthor{\bsnm{Casella},~\bfnm{George}\binits{G.}} \AND
\bauthor{\bsnm{McCulloch},~\bfnm{Charles~E.}\binits{C.~E.}}
(\byear{2006}).
\btitle{Variance Components}.
\bpublisher{Wiley}, \blocation{Hoboken, NJ}.
\bid{mr={2298115}}
\bptok{imsref}%
\end{bbook}
%
\endbibitem

%b33 #&#
\bibitem[\protect\citeauthoryear{Stewart and Sun}{1990}]{StewartSun1990}
%
\begin{bbook}[mr]
\bauthor{\bsnm{Stewart},~\bfnm{G.~W.}\binits{G.~W.}} \AND
\bauthor{\bsnm{Sun},~\bfnm{Ji~Guang}\binits{J.~G.}}
(\byear{1990}).
\btitle{Matrix Perturbation Theory}.
\bpublisher{Academic Press}, \blocation{Boston, MA}.
\bid{mr={1061154}}
\bptok{imsref}%
\end{bbook}
%
\endbibitem

%b34 #&#
\bibitem[\protect\citeauthoryear{Stinchcombe et~al.}{2010}]{Stinchcombe2010}
%
\begin{barticle}[pbm]
\bauthor{\bsnm{Stinchcombe},~\bfnm{John~R.}\binits{J.~R.}},
\bauthor{\bsnm{Izem},~\bfnm{Rima}\binits{R.}},
\bauthor{\bsnm{Heschel},~\bfnm{M.~Shane}\binits{M.~S.}},
\bauthor{\bsnm{McGoey},~\bfnm{Brechann~V.}\binits{B.~V.}} \AND
\bauthor{\bsnm{Schmitt},~\bfnm{Johanna}\binits{J.}}
(\byear{2010}).
\btitle{Across-environment genetic correlations and the frequency of selective
environments shape the evolutionary dynamics of growth rate in Impatiens
capensis}.
\bjournal{Evolution}
\bvolume{64}
\bpages{2887--2903}.
\bid{doi={10.1111/j.1558-5646.2010.01060.x}, issn={1558-5646}, pii={EVO1060},
pmid={20662920}}
\bptok{imsref}%
\end{barticle}
%
\endbibitem

%b35 #&#
\bibitem[\protect\citeauthoryear{Tibshirani}{1996}]{Tibshirani1996}
%
\begin{barticle}[mr]
\bauthor{\bsnm{Tibshirani},~\bfnm{Robert}\binits{R.}}
(\byear{1996}).
\btitle{Regression shrinkage and selection via the lasso}.
\bjournal{J. R. Stat. Soc. Ser. B Stat. Methodol.}
\bvolume{58}
\bpages{267--288}.
\bid{issn={0035-9246}, mr={1379242}}
\bptok{imsref}%
\end{barticle}
%
\endbibitem

\end{thebibliography}
\end{document}